\documentclass[prd,floatfix,amsmath,amssymb,10pt,nofootinbib,twocolumn]{revtex4}
\usepackage{amsmath,amssymb}
\usepackage{graphicx}
\usepackage{epstopdf}
\usepackage{color}
\usepackage{slashed}
\usepackage{wasysym,stmaryrd}
\usepackage{multirow}
\usepackage{array}
\usepackage[absolute]{textpos}
\usepackage{tikz}
\usetikzlibrary{arrows}
\usetikzlibrary{decorations.markings}

\usepackage{hyperref} % has to be last package
\usepackage[all]{hypcap} %references point to top of figures rather than captions
\usepackage{mciteplus}

\newcommand\ifb{{\ensuremath{\rm fb}^{-1}}}
\newcommand\lhc{{LHC}}
\newcommand{\comCL}[1]{{\color{red}\textsf{\small [#1]}}}

\newcommand{\hide}[1]{{}}

\newcommand{\mysection}[1]{{ \vspace{5mm} \section{#1} \vspace{-3mm} }}
\newcommand{\mystarsection}[1]{{ \vspace{5mm} \section*{#1} \vspace{-3mm} }}

\hide{
\newcommand{\muon}{pachydermion}
\newcommand{\Muon}{Pachydermion}
\newcommand{\muonic}{pachydermionic}
\newcommand{\muonSuffix}{{\rm p}} 
\newcommand{\brane}{{pachybrane}}

}

\newcommand{\muon}{heffalon}
\newcommand{\Muon}{Heffalon}
\newcommand{\muonic}{heffalonic}
\newcommand{\muonSuffix}{{\rm h}} 
\newcommand{\brane}{{heffalbrane}}

\hide{
\newcommand{\muon}{mastodon}
\newcommand{\Muon}{Mastodon}
\newcommand{\muonic}{mastodonic}
\newcommand{\muonSuffix}{{\rm ms}} 
\newcommand{\brane}{{mastobrane}}

}

\hide{
\newcommand{\muon}{proboscidon}
\newcommand{\Muon}{Proboscidon}
\newcommand{\muonic}{proboscidonean}
\newcommand{\muonSuffix}{{\rm pb}} 
\newcommand{\brane}{{proboscobrane}}

}

\newcommand{\muonCrossSection}{{\sigma_\muonSuffix}} 
\newcommand{\mmuon}{{{\bar M}_\muonSuffix}}

\newcommand{\ourxslimit}{{\ensuremath{64\,{\rm ab}}}}

\begin{document}
%\begin{textblock}{4cm}(0,0)
%1st April 2013
%\end{textblock}
%\date{1st April 2013}
\title{A search for direct \muon\ production using the ATLAS and CMS experiments at the Large Hadron Collider}

\author{Alan~J.~Barr}
\affiliation{Department of Physics, Denys Wilkinson Building, Keble Road, Oxford OX1 3RH, UK}
\author{Christopher~G.~Lester}
\affiliation{Department of Physics, Cavendish Laboratory, JJ Thomson
Avenue, Cambridge, CB3 0HE, UK}

\begin{abstract}
The first search is reported for direct \muon\ production, using 23.3~$\ifb$ per experiment of delivered integrated luminosity of proton-proton collisions at $\sqrt{s}=8$~TeV from the \lhc. The data were recorded with the ATLAS and the CMS detectors. Each exotic composite is assumed to be stable on the detector lifetime ($\tau$~$\gg$~ns). A particularly striking signature is expected. No signal events are observed after event selection. The cross section for \muon\ production is found to be less than \ourxslimit{} at the 95\% confidence level. 
\end{abstract}

\maketitle 

\textblockorigin{160mm}{10mm}
\begin{textblock}{300}[1,1](0,0)
\textsf{\large \color{gray}Submitted:~1~April~2013}
\end{textblock}

\mysection{Introduction}

The high energy of the Large Hadron Collider (LHC), together with the large integrated luminosity delivered to the general purpose experiments, 
offers unprecedented opportunities for searches for new  
particles beyond the Standard Model.
Given the wealth of analyses and searches that have been reported, 
it is perhaps surprising that to date
no search has been performed for some of the more exotic composites proposed in the literature.
%A variety of such heavy composites have been proposed.
In this paper we concentrate on the search for the \muon, 
for which the \lhc\ signature is expected to be particularly distinctive.

The \muon\ is a very heavy exotic composite particle~\cite{acacus,*job40:15,*deBelloAfrico} satisfying an approximate $Z_2$ symmetry. 
While \muonic\ models differ in their details~\cite{Saxe,*Kipling,*Dumbo,*Babar}
all models suggest a very large natural mass scale $\mmuon$, at around
$4\times 10^{30}$ GeV$/c^2$. 
The particle is expected to have a lifetime intermediate between that of the 
collider-detector scale ($\sim$ns) and the cosmological scale ($\sim$Gyr). 
%There are suggestions for the existence the particle even in the early literature (see for example Ref.~\cite{acacus,*sumer,*job40:15,*deBelloAfrico}), however it is regrettable that the literature contains significant inconsistencies
%~\cite{Babar,*Dumbo,*Kipling}. 

Cosmologically generated \muon s could have been generated in the hot, dense conditions in the early universe, however they have not yet been observed in either astronomical or cosmic ray experiments. Similarly no characteristic signature has so far been observed from annihilations in the galactic centre.\footnote{More details on the non-collider phenomenology are discussed in Section~\ref{sec:cosmics}.} 

In the absence of any convincing astronomical or cosmological evidence, an LHC search is both timely and appropriate.
Searches at other colliders~\cite{Aubert:2001tu} as well as other direct detection experiments~\cite{heffalumpSearchI,*heffalumpSearchII} have all so far failed to find any evidence 
supporting the \muon{} theory. 
%AJB removed this while he hunted for a natural place for it to fit in... 
%While some authors have found strong evidence for flaws \cite{Hovarth2012} in \muonic\ kinetic theory. 
As long as they continue to evade detection 
in collider experiments,
the existence of \muon s must be considered speculative.
%%%% Removed this footnote since paras 1, 2 and 3 of intro are now so good (and serious) that this weaker footnote gives the spoof game away too easily.  No real paper would say whether the PDF includes the particle name ... \footnote{Note, for example, that it has not yet been included among the particles listed by the Particle Data Group~\cite{Beringer:1900zz}.}

The production of \muon s at the LHC has a striking signature: \muon s can be expected to leave significant deposits in all detector components \cite{sumer}, even for near-threshold production. The large tracks would be distinctive and should have essentially no Standard Model background. Indeed one of the most identifiable effects of their transit through the inner detector is expected to be the characteristic loss of signal in subsequent bunch crossings over large portions of the detector volume.

We emphasise that while our results will be illustrated in terms of a particular simplified \muon\ model, essentially the same analysis would apply to other exotic composites with similar masses and interaction cross sections.
%In the discussion we identify other models to which similar limits would also apply.

\mysection{Detector and data samples}\label{sec:samples}

%\comCL{Regarding \MadeUpCollab, I understand the need not to void our membership of the collaboration, and to work within the ATLAS publications regulations which (as we know) extend even to papers outside the collaboration authored by ATLAS members. But if we can PROPERLY base our conclusions on REAL publicly available data (such as the luminosity taken by thereal ATLAS and CMS over time) then we can extract REAL limits on pink-elephant production, which leaves us potentially able to have the paper as a real one, eg Ignobel Prize material, even though it's signals are strange and the theory is improbable.  We can say ``The CMS and ATLAS'' experiments are ...''.  Alternatively, if we {\bf have}to make up a fictional detector, the option of ``The AtLAST-Collaboration'' also exists,  which might be better from a make-use-of-brain-dyslexia point of view. }

The ATLAS and CMS experiments~\cite{Aad:2008zzm,Chatrchyan:2008aa} are each jumbo-sized multi-purpose particle physics detectors,  
each with a forward-backward symmetric cylindrical
geometry and nearly 4$\pi$ coverage in solid angle.\footnote{It is of central importance to what follows that the reader appreciate that each 
experiment uses a right-handed coordinate system with its origin at the 
interaction point in the centre of the detector and the $z$-axis along the
beam pipe. Cylindrical coordinates $(r,\phi)$ are used in the transverse
plane, $\phi$ being the azimuthal angle around the beam pipe. The pseudorapidity $\eta$ is
defined in terms of the polar angle $\theta$ by $\eta=-\ln\tan(\theta/2)$, 
and the transverse energy $E_{\rm T}$ by $E_{\rm T}=E\sin\theta$.
} 
The detectors have the usual array of pixel detectors, silicon strip detectors, 
large superconducting magnets, calorimeters 
and muon chambers.
Since it is most unlikely that you are reading this section we 
leave the finer details to the imagination of the reader.
%experimental papers~\cite{Aad:2008zzm,Chatrchyan:2008aa}, 
%or if you prefer to the imagination of the reader.  

We note that the passage of a large-volume \muon\ composite
through either detector would generate significant structural deformations
of the detector system. The deformations expected would be much larger than those
observed during thermal cycling, so are very likely to be 
measurable using the detectors' precision laser alignment systems.
No attempt is made to reconstruct these `deformation' signatures,
which are beyond the scope of this paper,
however we encourage the experimental collaborations to investigate the 
feasibility of identifying such signals in future dedicated analyses.

The data samples used in this analysis were taken during
the period from March to December 2012
with the \lhc\ operating at a proton-proton centre-of-mass energy of $\sqrt{s}=8$~TeV.
It is the delivered luminosity rather than the recorded luminosity which is
the more important quantity in this analysis;
approximately 23.3\,fb$^{-1}$ of integrated luminosity 
was delivered to each experiment.

\mysection{\muon\ production at the LHC} 

As the natural \muonic\ mass scale $\mmuon$ is presumed to be around
$4\times 10^{30}$ GeV$/c^2$ (7 tonnes), one might expect direct
production of \muon s to be beyond the kinematic range of the
\lhc. However, the standard minimal \muonic\ model evades this
constraint by using the available centre-of-mass energy of the LHC only as
a trigger to initiate the {\em transport} of pre-existing \muon s from
a neighbouring brane onto ours in an extra-dimensional model.

More specifically, space-time is assumed to have $N$ small and compact
extra space-like dimensions, in addition to the usual four.  While all
Standard Model particles are confined to a (3+1)-dimensional
sub-manifold (the ``SM-brane'') \muon s and gravitons are, at least in
principle, able to exist at any point in the space.  In practice,
however, \muon s are predominantly expected to be found in only two
places: either in the vicinity of the SM-brane itself, or on another
sub-manifold (known as the ``\brane'') lying parallel to the SM-brane
but displaced by a very small length scale $\sim \delta$ in one or
more of the extra dimensions.
%In low-energy effective \muon\ theory, 
The source of the confinement of the \muon s on these two branes is
modelled with an effective potential having two degenerate minima
(one located on each brane) separated by a potential barrier. It is
the momentary lowering of this barrier caused by interactions
between it and gravitons produced in the primary $pp$ collision
that leads to the diffusion of \muon s from the \brane\ onto the
SM-brane and into the LHC detectors.

In short, \muon\ models assert that there is a \muon\ in the
room, in keeping with ideas already present in the literature of the
1930s.\footnote{ ``It is going beyond observation to assert there is
not an elephant in the room, for I cannot observe what is not''
\cite{KallenAndHook}.}  %The LHC is only needed to make it visible.

\mysection{{{Compatibility with dark matter and cosmic-ray data}
\label{sec:cosmics}}}

\Muon s have long been known to solve the dark-matter problem,\footnote{The
population of \muon s trapped on the \brane\ are known as the
dark-\muon s. The minimal hefflationary model predicts a mass
density of dark-\muon{} at $T_{\rm crit}$ (the temperature at which the
universe was last transparent to \muon s) in agreement with the recent
results from Planck~\cite{2013arXiv1303.5062P}. This solution of the dark
matter problem is one of the strongest reasons in support of this
class of \muonic\ models.} but the long standing deficit of atmospheric
\muon s in cosmic-ray data has caused concern for proponents of the
model.\footnote{The Tunguska incident notwithstanding.}  Nevertheless, it has been suggested that corrections from a quantum theory of gravity might cause the production of gravitons in collisions of SM particles to be suppressed by a warp $K$-factor
$$K(\xi,\Delta)\propto \exp\left[{-(C^\mu_{\
\nu\sigma\tau}C_\mu^{\ \nu\sigma\tau} -
\xi)^2/(2\Delta^2)}\right]$$
in which $C^\mu_{\
\nu\sigma\tau}$ is the Weyl-curvature of General Relativity.  The effect of a warp $K$-factor is thus to localise production of \muon s to regions of space in which the Weyl-curvature $C^2$ takes values close to the parameter $\xi$ and with a width controlled by $\Delta$.  How does the Weyl-curvature vary in our vicinity of our planet?   The solution of Einstein's equations for sphere of uniform density, mass $M$ and radius $R$, has a Weyl curvature equal to zero within the sphere, and sees it fall like $1/{r^6}$ outside it:\footnote{Here $r$ is the standard radial co-ordinate of the (exterior) Schwarzschild metric.  Note that the Weyl curvature outside a sphere of non-constant density, such as a real planet, would be smoothed out by density gradients.}

\vspace{4mm}
\begin{tikzpicture}[scale=2]

  \draw[->] (-0.2,0) -- (3.2,0) node[right] {$r$};
  \draw[->] (0,-0.2) -- (0,1.2) node[above] {$C^\mu_{\ \nu\sigma\tau}C_\mu^{\ \nu\sigma\tau}$};

  \foreach \x/\xtext in {1/R}
    \draw[shift={(\x,0)}] (0pt,2pt) -- (0pt,-2pt) node[below] {$\xtext$};

  \foreach \y/\ytext in {1/{\frac{48G^2M^2}{R^6}}}
    \draw[shift={(0,\y)}] (2pt,0pt) -- (-2pt,0pt) node[left] {$\ytext$};

%  \draw (-.5,.25) parabola bend (0,0) (2,4) node[below right] {$x^2$};
  %\draw (1,1) parabola bend (2,.25) (3,.125) node[below right] {$r^{-6}$};

  \draw plot file{lesterplot.dat};
  \draw (1.5,0.5) node {$\propto r^{-6}$};

 % \draw plot coordinates {(0,0) (1,1) (2,0) (3,1) (2,1) (10:2cm)};

\end{tikzpicture}

\noindent
Consequently, a value of $\xi$ near the natural value of $\xi_\text{crit}=48 G^2 M^2/R^6$ for the earth strongly suppresses \muon{} production in the upper atmosphere, on the surface of the moon, and on all cosmological objects having different values of $M^2/R^6$, while still allowing production to take place at the LHC.  Whether the fine-tuning required for $\xi$ and $\Delta$ is better or worse than that associated with the SM Higgs sector is the subject of on-going research.
\mysection{Simplified \muon\ models}
Limits on \muon s are thus usually expressed in the simplified-model space parametrised by three parameters $\{\muonCrossSection,T,C^2\}$. Here $\muonCrossSection$ is the direct \muon\ production cross section, $T$ is the temperature of the dark-\muon s, and $C^2$ is the square of the Weyl-curvature tensor.   Our current analysis is insensitive to $T$ since our counting experiment does not measure the \muon\ $p_T$ spectrum which $T$ controls.  Due to the location of the LHC we only present limits for $C^2 \approx 1.1 \times 10^{-10}~\text{s}^{-4}$. We can however place model-independent bounds on $\muonCrossSection$.

%\comAB{Do we need to address the heirarchy problem? Why is $\mmuon$ so much larger than the electroweak scale?}

%\comCL{No - the elephant mass scale is primarily related to the PLANK scale (1E20 GeV/csq) since elephants have to support themselves against gravity since they need to stay at a constant value of the Riemann Curvature Scalar (they don't move along geodesics).  alpha/electroweak sets the maximum compressibility of their bones, and this, in the limit they are 100\% bone, sets a limit on how big they can be -- basically thus $\mmuon$ can be written in terms of $\mplanck$ and alphaEW.  I'll do this.}

%\comCL{Brian Greene wrote `The Elegant Universe: Superstrings, Hidden Dimensions, and the Quest for the Ultimate Theory''.  Is there any way we can profit from talking about ``The {\em Elephant} Universe'' ?}

\hide{

\subsection{Bounds from other relevant experiments}
\begin{itemize}
\item
Tunguska incident.
 Relevant as we know that frozen mammoths are found in the same area, and the Weyl curvature is similar
\item
Dinosaur extinction event
\item
SN 1987A
\item
Crab Nebula supernova, 1054
\item
Elephant's trunk nebula
\item
Planck anomalies
\item
How can you tell if there's an elephant in your fridge --- there'll be footprints in the butter.
\end{itemize}

}

\hide{
\subsection{Future theoretical directions}

Should a signal be found, it will be important to distinguish
\muonic\ production from the microscopic black-hole production also
postulated in extra-dimensional models. Though this could prove to be
a mammoth task, close attention to the No-Hair Conjecture may
%\cite{Misner:1974qy}
may provide a way of discriminating one signal
from the other. \comCL{The No-Hair Conjecture asserts that all black
hole solutions of the Einstein-Maxwell equations of gravitation and
electromagnetism in general relativity can be completely characterized
by only three externally observable classical parameters: mass,
electric charge, and angular momentum.}

}

\mysection{Object and event selection}

\begin{figure}
\includegraphics[width=\columnwidth]{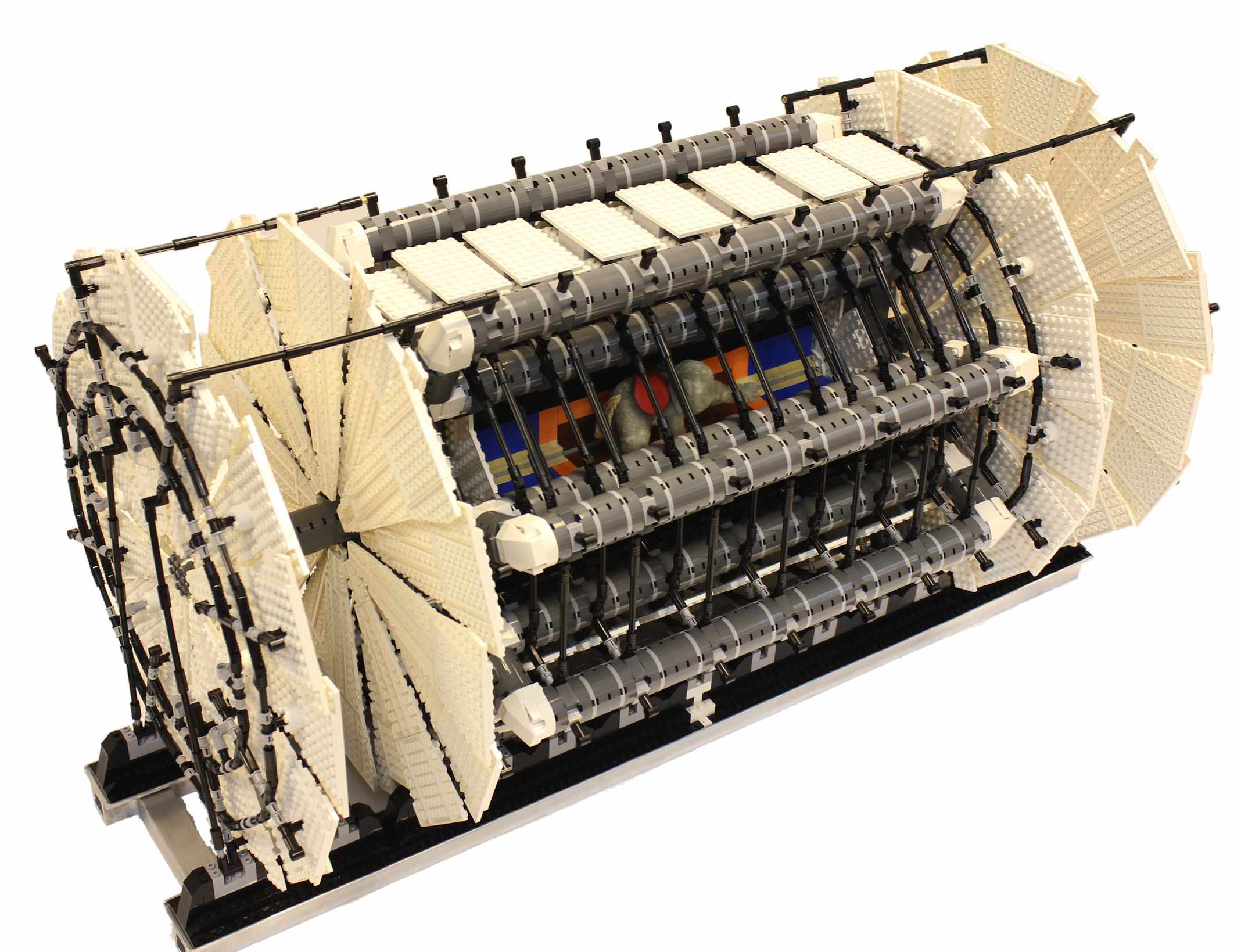}
\caption{\label{fig:display}
This simulated event illustrates a possible \muon\ production process within our toy model.
%Due to the limited data available about its structure, the precise
%appearance of the heavy composite is necessarily somewhat speculative.
The material deformation to the detector caused by the \muon\ transit has not been modelled.
}
\end{figure}

The characteristic signal for direct \muon\ production is a loss of 
signal from large areas of the detector in subsequent bunch crossings.
%Additional signals might be observed from large `ears' and perhaps even long tails in the
%the inner detector. However the detector evidence for such
%such secondary signals is likely to be difficult to reconstruct 
%(except for production very close to threshold), 
%and in any case could only be detected during the subsequent long shut-down period.
%Only the `smoking-gun' signature of signal loss is therefore considered in what follows.
The probability for even a low-mass \muon\ to tunnel 
through the entire ATLAS or CMS inner detector is small;
we therefore (conservatively) estimate the reconstruction efficiency to be $>$99\%.
The signature is largely unaffected by acceptance issues;
the probability to lose an entire \muon\ composite along the beam direction,
without any interaction whatever in the detection volume has been found to be negligible.
We further increase our confidence in the analysis method by 
independently searching for loss of signal in each of the separate sub-detector layers.
The event-to-event statistical correlations are fully taken into account.

Backgrounds from Standard Model processes have been modelled with the Pythia Monte Carlo program, 
and are found to be much smaller than one event.\footnote{The underlying event transverse energy distribution was found to differ from that given by the Pythia truth, therefore the data have been reweighted to the Monte Carlo.} 
Backgrounds from non-collision sources have been modelled using a robust data-driven approach, 
the details of which can be found in our supporting documentation
\cite{unavailble}.

%A display showing an artist's impression of a \muon{} candidate event can be found in 
A representation of a \muon{} candidate in the simplified detector simulation can be found in
Figure~\ref{fig:display}.

\mysection{Results and statistical analysis}

The {\em a priori} distribution of \muon\ masses and spins is subject to 
model uncertainties, hence we employ a robust frequentist analysis for which 
such subjective prejudices do not play a role. 
We emphasise that due to the innovative event selection and reconstruction methodology,
the analysis does not suffer from the effects of the nuisance parameters
(e.g. jet energy scale, resolution, etc) that typically plague other, less sophisticated, studies. 
Other than the statistical uncertainty due to the small event count,
the dominant residual  uncertainty results from the luminosity uncertainty (4\%). 

The number of events found after full object selection can be seen in 
Table~\ref{results}, together with the expected Standard Model and non-collision backgrounds.
No data-loss event is observed in any detector layer, 
hence it has not been possible to confirm the existence of the 
\muon.

\begin{table}[h]
\renewcommand\arraystretch{1.3}
\begin{tabular}{c|c|c|c|c}
\hline\hline
 & Sub-detector: & \ \ \ \  Inner \ \ \ \  & Calorimeter & \ \ \  Muon \ \ \ \\
\hline
\hline
\multirow{3}{*}{ATLAS\ } & Data & ${}^{\phantom{\dag}}$0$^\dag$ & 0 & 0 \\
\cline{2-5}
  & SM & 0 & 0 & 0 \\
\cline{2-5}
  & Non-collision & 0 & 0 & 0 \\
\hline\hline
\multirow{3}{*}{CMS} & Data & 0 & 0 & 0 \\
\cline{2-5}
  & SM & 0 & 0 & 0 \\
\cline{2-5}
  & Non-collision & 0 & 0 & 0 \\
\hline\hline
\end{tabular}
\caption{\label{results}
The number of characteristic pathological data-loss events recorded in each layer of the ATLAS and CMS detectors.
($^\dag$ One candidate event was recorded, however it was subsequently
ascribed to a cable mapping error, and hence was removed from the analysis.)
}
\end{table}

In the absence of a signal, a model-independent limit is placed on direct 
\muon\ production. Assuming a Poisson production probability 
distribution, the upper limit on the production cross section 
is found to be $\muonCrossSection < \ourxslimit$ at the 95\% confidence level.
% Figure~\ref{} shows how this model-independent limit compares to the predictions of our simplified model.

\mysection{Conclusion and outlook}
The \muon\ is only one of a large number of complex composite particles yet to 
receive verification in the collider environment. 
While this particular search has not yet
been able to confirm the existence of the elusive \muon{}, 
it has been able to set stringent limits on its production cross section.
The forthcoming LHC energy increase is expected to further increase the \muonic{} cross section
and so hope remains that the question of the existence of the
\muon{} finally can be settled within the next few years.

\mystarsection{Acknowledgments}
We wish to thank Martin Worthington and Luuk Huitink for providing
insight and expert knowledge on the relevant early literature, Luke
Butcher and Mike Hobson for advice relating to features of the production mechanism,
and Thomas Gillam for everything else.

\bibliographystyle{apsrevM}
\bibliography{Pachydermion}

\ifx\mcitethebibliography\mciteundefinedmacro
\PackageError{apsrevM.bst}{mciteplus.sty has not been loaded}
{This bibstyle requires the use of the mciteplus package.}\fi
\begin{mcitethebibliography}{16}
\expandafter\ifx\csname natexlab\endcsname\relax\def\natexlab#1{#1}\fi
\expandafter\ifx\csname bibnamefont\endcsname\relax
  \def\bibnamefont#1{#1}\fi
\expandafter\ifx\csname bibfnamefont\endcsname\relax
  \def\bibfnamefont#1{#1}\fi
\expandafter\ifx\csname citenamefont\endcsname\relax
  \def\citenamefont#1{#1}\fi
\expandafter\ifx\csname url\endcsname\relax
  \def\url#1{\texttt{#1}}\fi
\expandafter\ifx\csname urlprefix\endcsname\relax\def\urlprefix{URL }\fi
\providecommand{\bibinfo}[2]{#2}
\providecommand{\eprint}[2][]{\url{#2}}

\bibitem[{aca()}]{acacus}
\bibinfo{note}{A prehistoric elephant description in cave paintings from the
  Libian Tadrart Acacus, 12,000 B.C. :
  doi:10.1371/journal.pone.0049786.g002}\relax
\mciteBstWouldAddEndPuncttrue
\mciteSetBstMidEndSepPunct{\mcitedefaultmidpunct}
{\mcitedefaultendpunct}{\mcitedefaultseppunct}\relax
\EndOfBibitem
\bibitem[{job()}]{job40:15}
\bibinfo{howpublished}{The book of Job Ch. 40 vv 15-}\relax
\mciteBstWouldAddEndPuncttrue
\mciteSetBstMidEndSepPunct{\mcitedefaultmidpunct}
{\mcitedefaultendpunct}{\mcitedefaultseppunct}\relax
\EndOfBibitem
\bibitem[{deB(40 B.C.)}]{deBelloAfrico}
\emph{\bibinfo{title}{{De Bello Africo}}} (\bibinfo{year}{40 B.C.}),
  \bibinfo{note}{author unknown, commonly attributed to Julius Caesar}\relax
\mciteBstWouldAddEndPuncttrue
\mciteSetBstMidEndSepPunct{\mcitedefaultmidpunct}
{\mcitedefaultendpunct}{\mcitedefaultseppunct}\relax
\EndOfBibitem
\bibitem[{\citenamefont{Saxe}(1872)}]{Saxe}
\bibinfo{author}{\bibfnamefont{J.~G.} \bibnamefont{Saxe}},
  \emph{\bibinfo{title}{The blind men and the elephant}}
  (\bibinfo{year}{1872})\relax
\mciteBstWouldAddEndPuncttrue
\mciteSetBstMidEndSepPunct{\mcitedefaultmidpunct}
{\mcitedefaultendpunct}{\mcitedefaultseppunct}\relax
\EndOfBibitem
\bibitem[{\citenamefont{Kipling}(1902)}]{Kipling}
\bibinfo{author}{\bibfnamefont{R.}~\bibnamefont{Kipling}},
  \emph{\bibinfo{title}{Just so stories}} (\bibinfo{year}{1902})\relax
\mciteBstWouldAddEndPuncttrue
\mciteSetBstMidEndSepPunct{\mcitedefaultmidpunct}
{\mcitedefaultendpunct}{\mcitedefaultseppunct}\relax
\EndOfBibitem
\bibitem[{\citenamefont{Aberson and Pearl}(1941)}]{Dumbo}
\bibinfo{author}{\bibfnamefont{H.}~\bibnamefont{Aberson}} \bibnamefont{and}
  \bibinfo{author}{\bibfnamefont{H.}~\bibnamefont{Pearl}},
  \emph{\bibinfo{title}{Dumbo the Flying Elephant}}
  (\bibinfo{publisher}{Whitman}, \bibinfo{year}{1941})\relax
\mciteBstWouldAddEndPuncttrue
\mciteSetBstMidEndSepPunct{\mcitedefaultmidpunct}
{\mcitedefaultendpunct}{\mcitedefaultseppunct}\relax
\EndOfBibitem
\bibitem[{\citenamefont{de~Brunhof}(1976)}]{Babar}
\bibinfo{author}{\bibfnamefont{J.}~\bibnamefont{de~Brunhof}},
  \emph{\bibinfo{title}{Histoire de Babar}} (\bibinfo{year}{1976})\relax
\mciteBstWouldAddEndPuncttrue
\mciteSetBstMidEndSepPunct{\mcitedefaultmidpunct}
{\mcitedefaultendpunct}{\mcitedefaultseppunct}\relax
\EndOfBibitem
\bibitem[{\citenamefont{Aubert et~al.}(2002)}]{Aubert:2001tu}
\bibinfo{author}{\bibfnamefont{B.}~\bibnamefont{Aubert}} \bibnamefont{et~al.}
  (\bibinfo{collaboration}{{BABAR}}), \bibinfo{journal}{Nucl.Instrum.Meth.}
  \textbf{\bibinfo{volume}{A479}}, \bibinfo{pages}{1} (\bibinfo{year}{2002}),
  \eprint{hep-ex/0105044}\relax
\mciteBstWouldAddEndPuncttrue
\mciteSetBstMidEndSepPunct{\mcitedefaultmidpunct}
{\mcitedefaultendpunct}{\mcitedefaultseppunct}\relax
\EndOfBibitem
\bibitem[{\citenamefont{Robin et~al.}(1926)\citenamefont{Robin, Pooh, and
  Piglet}}]{heffalumpSearchI}
\bibinfo{author}{\bibfnamefont{C.}~\bibnamefont{Robin}},
  \bibinfo{author}{\bibfnamefont{W.~T.} \bibnamefont{Pooh}}, \bibnamefont{and}
  \bibinfo{author}{\bibnamefont{Piglet}}, \emph{\bibinfo{title}{Winnie the
  Pooh}} (\bibinfo{publisher}{Methuen \& Co.~Ltd.}, \bibinfo{year}{1926}),
  chap. \bibinfo{chapter}{5: ``In which Piglet meets a Heffalump''}\relax
\mciteBstWouldAddEndPuncttrue
\mciteSetBstMidEndSepPunct{\mcitedefaultmidpunct}
{\mcitedefaultendpunct}{\mcitedefaultseppunct}\relax
\EndOfBibitem
\bibitem[{\citenamefont{Robin et~al.}(1928)\citenamefont{Robin, Pooh, and
  Piglet}}]{heffalumpSearchII}
\bibinfo{author}{\bibfnamefont{C.}~\bibnamefont{Robin}},
  \bibinfo{author}{\bibfnamefont{W.~T.} \bibnamefont{Pooh}}, \bibnamefont{and}
  \bibinfo{author}{\bibnamefont{Piglet}}, \emph{\bibinfo{title}{The House at
  Pooh Corner}} (\bibinfo{publisher}{Methuen \& Co.~Ltd.},
  \bibinfo{year}{1928}), chap. \bibinfo{chapter}{3: ``In which a search is
  organised, and Piglet nearly meets the Heffalump again''}\relax
\mciteBstWouldAddEndPuncttrue
\mciteSetBstMidEndSepPunct{\mcitedefaultmidpunct}
{\mcitedefaultendpunct}{\mcitedefaultseppunct}\relax
\EndOfBibitem
\bibitem[{\citenamefont{Alster}(1997)}]{sumer}
\bibinfo{author}{\bibfnamefont{B.}~\bibnamefont{Alster}},
  \emph{\bibinfo{title}{Proverbs of ancient Sumer : the world's earliest
  proverb collection}} (\bibinfo{publisher}{CDL Press}, \bibinfo{year}{1997}),
  ISBN \bibinfo{isbn}{188305320X}, \bibinfo{note}{proverb 121 5.1 : ``{\it umma
  pi-ru-um ina ramani\v sa ina b\= ul \v Samkan \v sa k\=\i ma j\^ atima z\^ u
  ul iba\v s\v si}'' [... the elephant said to herself, `Among the wild
  creatures of {\v S}akan, there is no one who can defecate like me']
  (1900-1600 B.C.).}\relax
\mciteBstWouldAddEndPunctfalse
\mciteSetBstMidEndSepPunct{\mcitedefaultmidpunct}
{}{\mcitedefaultseppunct}\relax
\EndOfBibitem
\bibitem[{\citenamefont{Aad et~al.}(2008)}]{Aad:2008zzm}
\bibinfo{author}{\bibfnamefont{G.}~\bibnamefont{Aad}} \bibnamefont{et~al.}
  (\bibinfo{collaboration}{ATLAS}), \bibinfo{journal}{JINST}
  \textbf{\bibinfo{volume}{3}}, \bibinfo{pages}{S08003}
  (\bibinfo{year}{2008})\relax
\mciteBstWouldAddEndPuncttrue
\mciteSetBstMidEndSepPunct{\mcitedefaultmidpunct}
{\mcitedefaultendpunct}{\mcitedefaultseppunct}\relax
\EndOfBibitem
\bibitem[{\citenamefont{Chatrchyan et~al.}(2008)}]{Chatrchyan:2008aa}
\bibinfo{author}{\bibfnamefont{S.}~\bibnamefont{Chatrchyan}}
  \bibnamefont{et~al.} (\bibinfo{collaboration}{CMS}), \bibinfo{journal}{JINST}
  \textbf{\bibinfo{volume}{3}}, \bibinfo{pages}{S08004}
  (\bibinfo{year}{2008})\relax
\mciteBstWouldAddEndPuncttrue
\mciteSetBstMidEndSepPunct{\mcitedefaultmidpunct}
{\mcitedefaultendpunct}{\mcitedefaultseppunct}\relax
\EndOfBibitem
\bibitem[{\citenamefont{Kallen and Hook}(1935)}]{KallenAndHook}
\bibinfo{author}{\bibfnamefont{H.~M.} \bibnamefont{Kallen}} \bibnamefont{and}
  \bibinfo{author}{\bibfnamefont{S.}~\bibnamefont{Hook}},
  \emph{\bibinfo{title}{American philosophy today and tomorrow}}
  (\bibinfo{publisher}{New York: L. Furman, inc}, \bibinfo{year}{1935})\relax
\mciteBstWouldAddEndPuncttrue
\mciteSetBstMidEndSepPunct{\mcitedefaultmidpunct}
{\mcitedefaultendpunct}{\mcitedefaultseppunct}\relax
\EndOfBibitem
\bibitem[{\citenamefont{{Ade} et~al.}(2013)}]{2013arXiv1303.5062P}
\bibinfo{author}{\bibfnamefont{P.~A.~R.} \bibnamefont{{Ade}}}
  \bibnamefont{et~al.} (\bibinfo{collaboration}{{Planck}}),
  \bibinfo{journal}{arXiv e-prints}  (\bibinfo{year}{2013}),
  \eprint{1303.5062}\relax
\mciteBstWouldAddEndPuncttrue
\mciteSetBstMidEndSepPunct{\mcitedefaultmidpunct}
{\mcitedefaultendpunct}{\mcitedefaultseppunct}\relax
\EndOfBibitem
\bibitem[{una()}]{unavailble}
\bibinfo{howpublished}{\url{http://secret_documents.cds.cern.ch}}\relax
\mciteBstWouldAddEndPuncttrue
\mciteSetBstMidEndSepPunct{\mcitedefaultmidpunct}
{\mcitedefaultendpunct}{\mcitedefaultseppunct}\relax
\EndOfBibitem
\end{mcitethebibliography}
\end{document}